\pdfoutput=1
\documentclass[twocolumn,preprintnumbers,superscriptaddress,aps]{revtex4-2}

\usepackage{amsthm,amsmath,amssymb,physics,mathtools}
\usepackage{bbm}

\usepackage{array,esint}
\def\CMD#1{%
   $ \csname#1\endcsname \displaystyle\csname#1\endcsname $ & \texttt{\textbackslash#1} &}

\usepackage{hyperref}
\hypersetup{
    colorlinks = true,       
    linkcolor = red,          
    citecolor = magenta,        
    filecolor=red,      
    urlcolor= cyan           
}

\usepackage{graphicx}
\graphicspath{ {TBA-PRL/} }
\usepackage[dvipsnames]{xcolor}
\definecolor{darkbrown}{rgb}{0.4, 0.26, 0.13}

\usepackage{dsfont}

\usepackage{tikz}
\usetikzlibrary{positioning,arrows}
\usepgflibrary{arrows.meta}
\usetikzlibrary{decorations.pathmorphing}
\usetikzlibrary{decorations.markings}
\usetikzlibrary{math}
\usetikzlibrary{calc}

\tikzset{
Wilson_1/.style={double distance=1.1pt,postaction={decorate}, decoration={markings,mark=at position .08 with {\arrow{Stealth[scale=0.5]}},mark=at position .985 with {\arrow{Stealth[scale=0.5]}}}},
Scalar/.style={densely dashed},
Gauge/.style={decorate, draw=black, decoration={coil,aspect=0.5, post length = 0pt, pre length = 0pt, segment length=1.75pt,amplitude=1.4pt}},
Gauge-Back/.style={ultra thick, opacity=0.8, decorate, draw=white, decoration={coil,aspect=0.5, post length = 0pt, pre length = 0pt, segment length=1.75pt,amplitude=1.4pt}},
Fermion/.style={},
Fermion-Back/.style={ultra thick, opacity=0.8, draw=white}, 
}

\usepackage{tikz}
\usepackage{pgf}
\usepackage{tcolorbox}
\usetikzlibrary{tikzmark}

\usetikzlibrary{decorations.pathmorphing}
\usetikzlibrary{intersections}
  \usetikzlibrary{positioning}
\usetikzlibrary{shapes}
\usetikzlibrary{fit}
\usetikzlibrary{backgrounds}

\usetikzlibrary{decorations.pathreplacing,calc,arrows,decorations.markings}

\newcommand{\be}{\begin{equation}}
\newcommand{\ee}{\end{equation}}
\newcommand{\beq}{\begin{equation}}
\newcommand{\eeq}{\end{equation}}
\newcommand{\bea}{\begin{eqnarray}}
\newcommand{\eea}{\end{eqnarray}}

\newcommand{\lu}{l}
\newcommand{\lv}{\ell}
\newcommand{\K}{\mathbb{K}}

\newcommand{\Gc}{\Gamma_{\textrm{cusp}}}

\newcommand{\N}{\mathcal{N}}

\newcommand{\Or}{\text{O}}

\newcommand{\G}{\mathcal{G}}

\newcommand{\A}{\mathcal{A}}
\newcommand{\cS}{\mathcal{S}}
\newcommand{\cP}{\mathcal{P}}

\newcommand{\la}{\langle}
\newcommand{\ra}{\rangle}
\newcommand{\ab}[1]{\la #1 \ra}

\begin{document}

\preprint{SLAC--PUB--17710}
\preprint{DESY-22-182}

\title{An Origin Story for Amplitudes}
\author{Benjamin~Basso}
\email{benjamin.basso@phys.ens.fr}
\affiliation{Laboratoire de Physique de l'Ecole Normale Sup\'erieure, ENS, Universit\'e PSL, CNRS, Sorbonne Universit\'e, Universit\'e Paris Cit\'e, F-75005 Paris, France}
\author{Lance~J.~Dixon}
\email{lance@slac.stanford.edu }
\affiliation{SLAC National Accelerator Laboratory, Stanford University, Stanford, CA 94309, USA}
\author{Yu-Ting~Liu}
\email{aytliu@stanford.edu }
\affiliation{SLAC National Accelerator Laboratory, Stanford University, Stanford, CA 94309, USA}
\affiliation{Kavli Institute for Theoretical Physics, UC Santa Barbara, Santa Barbara, CA 93106, USA}
\author{Georgios~Papathanasiou}
\email{georgios.papathanasiou@desy.de }
\affiliation{Deutsches Elektronen-Synchrotron DESY, Notkestr. 85, 22607 Hamburg, Germany}

\date{\today}

\begin{abstract}
We classify origin limits of maximally helicity violating
multi-gluon scattering amplitudes in planar $\N=4$ super-Yang-Mills theory,
where a large number of cross ratios approach zero,
with the help of cluster algebras. By analyzing existing perturbative data,
and bootstrapping new data, we provide evidence that the amplitudes become the
exponential of a quadratic polynomial in the large logarithms.
With additional input from the thermodynamic Bethe ansatz at strong coupling,
we conjecture exact expressions for amplitudes with up to 8 gluons in all origin limits.
Our expressions are governed by the tilted
cusp anomalous dimension evaluated at various values of the tilt angle.

\begin{center}
``Those who explore an unknown world are travelers without a map:
the map is the result of the exploration. The position of their
destination is not known to them, and the direct path that leads
to it is not yet made.'' {\it - Hideki Yukawa}
\end{center}
\end{abstract}

\maketitle

\section{Introduction}\label{sec:intro}
For generic kinematics, perturbative scattering amplitudes can be
extremely complicated functions of the kinematic variables.
In certain limits, they may simplify enormously.
For general gauge theories, simplifying kinematics include Sudakov regions,
where soft gluon radiation is suppressed, and high-energy or multi-Regge
limits, where Regge factorization holds.  In planar $\N=4$ super-Yang-Mills
theory (SYM), the duality of amplitudes to polygonal Wilson
loops~\cite{Alday:2007hr,Drummond:2007aua,Brandhuber:2007yx,Drummond:2007au}
allows near-collinear limits to be computed~\cite{Alday:2010ku,Basso:2013vsa}
in terms of excitations of the Gubser-Klebanov-Polyakov 
flux tube~\cite{Gubser:2002tv,Alday:2007mf}.
Recently, an even simpler kinematical region for six-gluon
scattering in the  maximally-helicity-violating (MHV) configuration
was found~\cite{Caron-Huot:2019vjl,Basso:2020xts},
the {\it origin} where all three cross ratios of the dual hexagon
Wilson loop are sent to zero.
In this limit, the logarithm of the MHV amplitude becomes
quadratic in the logarithms of the cross ratios.  The coefficients
of the two quadratic polynomials, $\Gamma_{\rm oct}$ and $\Gamma_{\rm hex}$,
can be computed for any value of the 't Hooft coupling
$\lambda \equiv g^2/(16\pi^2)$ by deforming
the Beisert-Eden-Staudacher (BES) kernel~\cite{Beisert:2006ez} by a
{\it tilt} angle $\alpha$, giving rise to a
``tilted cusp anomalous dimension''
$\Gamma_\alpha(g)$ (see eq.~(\ref{eq:Gamma_alpha})).
The usual BES kernel and cusp anomalous dimension
are recovered by setting $\alpha = \pi/4$,
$\Gamma_{\rm cusp} = \Gamma_{\alpha = \pi/4}$,
while the two hexagon-origin coefficients are given by
$\Gamma_{\rm oct} = \Gamma_{\alpha=0}$ and
$\Gamma_{\rm hex} = \Gamma_{\alpha=\pi/3}$.

This Letter will explore analogous origins for higher-point
MHV amplitudes, regions where the same quadratic logarithmic (QL)
behavior holds.
We will see that there is a cornucopia of such regions at seven and
especially eight points.  The regions need not be isolated points;
they can be one-dimensional lines starting at seven points, and up
to three-dimensional surfaces starting at eight points.
They can be classified by cluster algebras~\cite{1021.16017,1054.17024}, which provide natural compactifications of the space of positive kinematics~\cite{FG03b,Arkani-Hamed:2012zlh,Golden:2013xva,Fock2016ClusterPV}, at the boundary of which these limits are located.
Furthermore, we will provide a master formula that we conjecture
organizes the QL behavior of MHV amplitudes in all of these regions for
arbitrary coupling,
as a discrete sum over tilt angles,
in which $\Gamma_\alpha(g)$ carries all of the coupling dependence.
Our formula is motivated by
studying the thermodynamic Bethe ansatz (TBA)
representation~\cite{Alday:2009dv,Alday:2010vh,Alday:2010ku,Bonini:2015lfr}
of the minimal-area formula~\cite{Alday:2007hr}
for the amplitude at strong coupling.

\section{Classifying Origin Limits}
Dual conformal symmetry~\cite{Drummond:2006rz,Alday:2007hr,Drummond:2007aua,Brandhuber:2007yx,Drummond:2007au} in planar $\N=4$ SYM implies that
MHV amplitudes for $n$ gluons depend on $3(n-5)$ independent kinematical
variables. These may be chosen as a subset of the
$n(n-5)/2$ dual conformal cross ratios,
\be \label{eq:u_def}
u_{i,j} = \frac{x_{i,j+1}^2 x_{j,i+1}^2}{x_{i,j}^2 x_{j+1,i+1}^2} \,,
\ee
where $x_{i,j}^\mu \equiv p_i^\mu + p_{i+1}^\mu + \cdots + p_{j-1}^\mu$
are sums of cyclicly adjacent gluon momenta, and indices
are always mod $n$.

For $n=6$, all cross ratios $u_i \equiv u_{i+1,i+4}$, $i=1,2,3$ are independent,
and the origin limit is simply defined as the kinematic point
\be \label{eq:O6_def}
\Or^{(6)}:\ \ u_{i}\to0\,,\quad i=1,2,3.
\ee
At higher $n$, there are $(n-5)(n-6)/2$ Gram determinant polynomial relations
between the cross ratios, because there are a limited number of independent
vectors in fixed spacetime dimensions.  (For their explicit form for $n=7,8$,
see appendix~\ref{appx:Gramdet}.)
These relations raise the question of how to define the appropriate
generalizations of the origin limit.

To answer this question, we consider the \emph{positive region}, a
subregion of Euclidean scattering kinematics where amplitudes are expected to be devoid of branch points~\cite{Arkani-Hamed:2012zlh,Arkani-Hamed:2019rds}.  Thus the first
place to look for simple divergent behavior is at pointlike limits
at the boundary of the positive region.  Such limits may be
found systematically using \emph{cluster algebras}~\cite{1021.16017,1054.17024} associated with the
Grassmannian ${\rm Gr}(4,n)$~\cite{scott}, which provide a compactification of the positive region~\cite{FG03b,Arkani-Hamed:2012zlh,Golden:2013xva,Fock2016ClusterPV}, see also~\cite{Papathanasiou:2022lan}.
Accordingly, the positive region may be mapped to the inside of a polytope,
whose boundary comprises vertices connected by edges that bound polygonal faces,
that bound higher-dimensional polyhedra.
Cluster algebras, or more precisely cluster Poisson varieties,
consist of a collection of clusters, each containing $3(n-5)$ cluster
$\mathcal{X}$-coordinates $\mathcal{X}_i$, corresponding to a
coordinate chart describing this compactification.
Setting all $\mathcal{X}_i\to 0$ yields a vertex at the
boundary of the positive region. Letting all but one
$\mathcal{X}_i$ vanish gives an edge connecting
neighboring clusters, known as a \emph{mutation}. It is also associated with a birational transformation between the $\mathcal{X}$-coordinates of the connected clusters, enabling the generation of a cluster algebra from an initial cluster.

\begin{table}[h]
\centering
\begin{tabular}[t]{| l | c c c c c c c c | c c c c | c |}
\hline
Origin Class
& $u_1$ & $u_2$ & $u_3$ & $u_4$ & $u_5$ & $u_6$ & $u_7$ & $u_8$ & $v_1$ & $v_2$ & $v_3$ & $v_4$\\
\hline\hline
$\Or_1 (\textrm{super})$
& 0 & 0 & 0 & 0 & 0 & 0 & 0 & 0 & 0 & 1 & 0 & 1 \\\hline
$\Or_2$
& 0 & 0 & 0 & 0 & 0 & 0 & 0 & 1 & 0 & 1 & 0 & 1 \\\hline
$\Or_3$
& 0 & 0 & 0 & 0 & 0 & 0 & 0 & 1 & 0 & 0 & 1 & 1 \\\hline
$\Or_4$
& 0 & 0 & 0 & 0 & 0 & 0 & 1 & 1 & 0 & 0 & 1 & 0 \\\hline
$\Or_5$
& 0 & 0 & 0 & 0 & 0 & 1 & 0 & 1 & 0 & 0 & 1 & 0 \\\hline
$\Or_6$
& 0 & 0 & 0 & 0 & 1 & 0 & 0 & 1 & 0 & 1 & 0 & 0 \\\hline
$\Or_7$
& 0 & 0 & 0 & 0 & 1 & 0 & 0 & 1 & 0 & 0 & 1 & 0 \\\hline
$\Or_8$
& 0 & 0 & 0 & 1 & 0 & 0 & 0 & 1 & 0 & 1 & 0 & 0 \\\hline
$\Or_9$
& 0 & 0 & 0 & 1 & 0 & 0 & 0 & 1 & 0 & 0 & 0 & 1 \\\hline\end{tabular}
\caption{All dihedrally inequivalent origin classes for $n=8$.
Zeros represent infinitesimal values. There are nine infinitesimal
cross ratios for all origins except for the \emph{super-origin} $\Or_1$
which has ten. All nonzero cross ratios are close to unity.}
\label{tab:octorigins}
\end{table}

We start with the finite ${\rm Gr}(4,n)$ cluster algebras for $n=6,7$,
with Dynkin labels $A_3$ and $E_6$~\cite{1054.17024}.
We first observe that in all boundary vertices, $u_{i,j} =0$~or~$1$.
These kinematic points contain the $n=6$ origin limit~(\ref{eq:O6_def});
at $n=7$ we find 28 clusters describing analogous limits where all
but one of the seven $u_i \equiv u_{i+1,i+4}$, $i=1,2,\ldots,7$, vanishes,
\be \label{eq:O7j_def}
\Or_j^{(7)}:\  u_{i\neq j}\to0\,,\quad u_j=1\,.
\ee
The seven origins are related by a cyclic symmetry, $u_i\mapsto u_{i+1}$.
There are four clusters for each $\Or_j^{(7)}$,
two with a different direction of approach to the limit,
plus their parity images.
All of these clusters form a cyclic chain connected by mutations
or lines in the space of kinematics.
In terms of cross ratios, the line connecting $\Or_7^{(7)}$ and $\Or_1^{(7)}$
is
\be \label{eq:Line71_def}
\hbox{\sc Line 71}:\ \ u_i\ll 1, \quad i=2,3,4,5,6; \quad u_7+u_1 = 1\,,
\ee
with $u_1,u_7 \in [0, 1]$. The remaining lines are obtained by cyclic symmetry.
Quite remarkably, the amplitude exhibits exponentiated QL behavior
not only on the points~\eqref{eq:O7j_def}, but also on these origin lines!
This QL behavior also implies that the value of the amplitude is
independent of the direction or speed of approach to the limit;
it remains the same function of the cross ratios irrespective
of the rate with which they tend to zero.

Inspired by these examples, we define \emph{origin points} at higher $n$ as vertices
where at least $3(n-5)$ cross ratios approach zero. We now
classify the $n=8$ origin points. While the corresponding ${\rm Gr}(4,8)$
cluster algebra is infinite-dimensional, there is a procedure for selecting
a finite subset of clusters~\cite{Drummond:2019qjk,Drummond:2019cxm,%
Arkani-Hamed:2019rds,Henke:2019hve,Henke:2021ity} based on
tropicalization~\cite{Speyer2005}, see also~\cite{Arkani-Hamed:2019mrd}.
Here we start with a cluster corresponding to an origin point,
and generate new clusters by mutations until this condition is no
longer met.
We find 1188 clusters contained in the finite subset selected
in~\cite{Drummond:2019qjk,Drummond:2019cxm,Arkani-Hamed:2019rds,%
Henke:2019hve,Henke:2021ity}, as further described in appendix~\ref{appx:kinematics} and in an ancillary file.
Modding out by parity, dihedral symmetry and direction of approach,
these origins belong to the nine classes shown in
table~\ref{tab:octorigins}, where $u_i \equiv u_{i+1,i+4}$, $i=1,2,\ldots,8$,
and $v_i \equiv u_{i+1,i+5}$, $i=1,2,3,4$.  This table may be obtained even
more simply by assuming that all cross ratios approach 0 or 1,
and scanning for all combinations that satisfy the Gram
determinant constraints.
This process also identifies one more potential origin,
$\Or_{\rm X}=(0,0,1,0,0,1,0,1;0,0,0,0)$ in the $(u_i;v_j)$ notation of
table~\ref{tab:octorigins}.  It lies outside of the positive region,
and we defer its study to future work.

\begin{figure}[t]
\begin{center}
\includegraphics[width=3.0in]{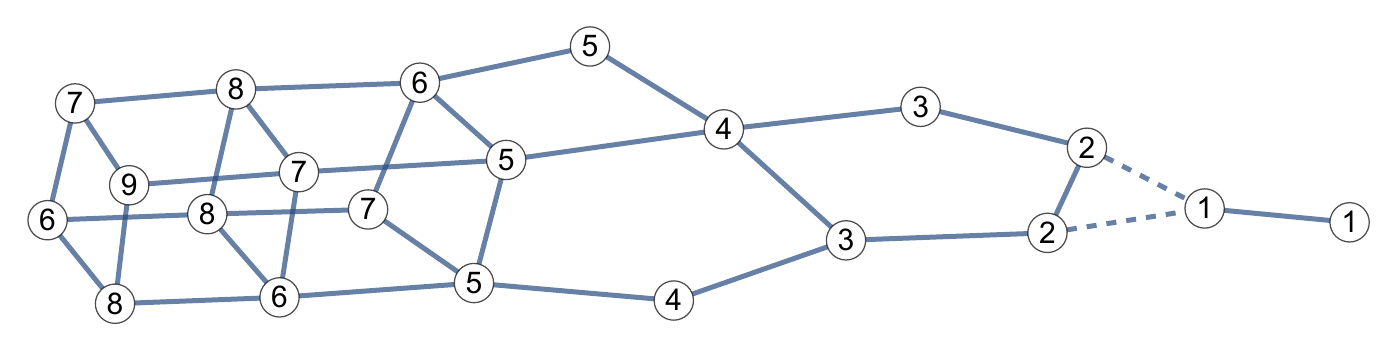} 
\end{center}
\caption{The system of eight-point origins exhibiting QL
behavior. We omit many origins that are related to the ones shown
by dihedral symmetry. The node numbers correspond to $\Or_i$ in
table~\ref{tab:octorigins} or their dihedral images.
The behavior on the lines and surfaces shown in the figure is
also QL, except for the dashed line between $\Or_1$
and $\Or_2$.}
\label{fig:octbdy}
\end{figure}

At $n=8$, there are also higher-dimensional QL surfaces connecting
the $\Or_i$, which generalize the seven-point
{\sc Line 71}~(\ref{eq:Line71_def}).
Motivated by this line, which also defines an $A_1$ subalgebra of the $E_6$ cluster algebra, we searched for maximal subalgebras of 
the ${\rm Gr}(4,8)$ cluster algebra that move one solely from origin to origin.
Two $A_3$ subalgebras correspond to two cubes,
{\sc Cube 6789} and {\sc Cube 5678}~\footnote{%
The $A_3$ polytope is a truncated triangular bipyramid, but if we collapse
its vertices that correspond to the same points in cross-ratio space,
we get a cube in both cases.}.
Two $A_2$ subalgebras correspond to {\sc Pentagon 345} and {\sc Pentagon 234}.
An $A_1\times A_1$ corresponds to {\sc Square 456}.
An $A_1$ subalgebra {\sc Superline 1} connects two super-origins $\Or_1$.
These high-dimensional spaces interpolating between origins are summarized
in table~\ref{tab:octoriginsurfacesrelation} and are depicted in
figure~\ref{fig:octbdy}.

\begin{table}[h]
\centering
\begin{tabular}[t]{| l | c |}
\hline
Boundary & Relations \\
\hline\hline
{\sc Cube 6789}
&  $u_3+u_4 = u_7+u_8 = v_1+v_4 = 1$ \\\hline
{\sc Cube 5678}
&  $u_1+u_2 = u_4+u_5 = v_2+v_3 = 1$ \\\hline
{\sc Square 456}
&  $u_1+u_2 = u_7+u_8 = v_4 = 1$ \\\hline
{\sc Pentagon 345}
&  $u_8+u_1u_7 = u_7+u_8v_4 = v_4+u_7v_3 = 1$ \\\hline
{\sc Pentagon 234}
&  $u_1+u_8v_3 = v_3+u_1v_4 = u_8+u_1v_1 = 1$ \\\hline
{\sc Superline 1}
&  $v_1=1-v_2=v_3=1-v_4$ \\\hline
\end{tabular}
\caption{Relations among the finite cross ratios for the octagon boundaries. All the cross ratios unspecified in the relations are implicitly infinitesimal.}
\label{tab:octoriginsurfacesrelation}
\end{table}

\section{Perturbative Data \& Bootstrap}
In this Letter we work with the $n$-point remainder function $R_n$, related to the MHV amplitude by
\be
\exp R_n\equiv {A_n^\text{MHV}}/{A_n^\text{BDS}}\,, \nonumber
\ee
where the known, infrared-divergent normalization factor $A_n^\text{BDS}$
is essentially the exponential of the one-loop
amplitude~\cite{Bern:2005iz,Bern:2008ap,Drummond:2008aq}.
The remainder function is infrared finite, and invariant under dual
conformal symmetry as well as the $n$-gon dihedral symmetry group $D_n$.

Using perturbative data through seven loops, $R_6$ was found to simplify drastically~\cite{Caron-Huot:2019vjl} at the origin~(\ref{eq:O6_def}): To $\mathcal{O}(u_i^0)$, it becomes the sum of two QL polynomials,
\be \label{eq:poly6}
R_6  {= -\tfrac{\Gamma_0-\Gamma_{\pi/4}}{24}\ln^{2}{(u_1 u_2 u_3)} -\tfrac{\Gamma_{\pi/3}-\Gamma_{\pi/4}}{24} \sum_{i=1}^{3} \ln^{2}{(\tfrac{u_i}{u_{i+1}})}
}\,,
\ee
where each polynomial is multiplied by the \emph{tilted}
cusp anomalous dimension $\Gamma_{\alpha}$ evaluated at different angles
$\alpha = 0,\tfrac{\pi}{4},\tfrac{\pi}{3}$~\cite{Basso:2020xts}.
For $n=6$, $D_6$ acts on the $u_i$ as arbitrary $S_3$ permutations.
The origin preserves this symmetry, so only $S_3$-symmetric
quadratic polynomials are allowed,
which are exhausted by those of eq.~(\ref{eq:poly6}).

For $n=7$, QL behavior was observed for $R_7$ through four loops
at the dihedrally-equivalent origins $\Or_j^{(7)}$~\cite{Dixon:2020cnr}.
More generally, a four-loop computation
along the lines of ref.~\cite{Dixon:2020cnr}
reveals that the remainder function $R_7$ on
{\sc Line 71}~(\ref{eq:Line71_def}) is given by,
\be \label{eq:R7_origin_line}
R_7({\hbox{\sc Line 71}}) = \sum_{i=1}^3 c_i P_i^{(7)} \,,
\ee
where
\begin{align} \label{eq:R7_Pi}
P_1^{(7)}
&= \sum_{i=1}^6 \lu_i\lu_{i+1} + \sum_{i=1}^5 \lu_i\lu_{i+2} \,, \nonumber\\
P_2^{(7)} &= -\lu_1\lu_7 + \sum_{i=1}^7 \lu_i^2 + \sum_{i=1}^4 \lu_i\lu_{i+3}
 \,, \nonumber\\
P_3^{(7)} &= \sum_{i=1}^7 \lu_i\lu_{i+2} - \sum_{i=1}^3 \lu_i \lu_{i+4} \,,
\end{align}
are quadratic polynomials in the logarithms, $\lu_i \equiv \ln u_i$.
In eq.~(\ref{eq:R7_origin_line}) and in the following, we give only
the leading QL behavior in the given limit.
We never find any linear-logarithmic terms. There are constant terms followed by subleading power
corrections, which we do not study.

Through four loops, the coefficients $c_i$ in eq.~(\ref{eq:R7_origin_line})
are given by,
\begin{align} \label{eq:R7_cs}
c_1 &= g^4 \zeta_2 - \tfrac{37}{2} g^6 \zeta_4 + g^8 \left(
    \tfrac{1975}{8} \zeta_6 - 2 \zeta_3^2 \right) + O(g^{10}) \,, \nonumber\\
c_2 &= -\tfrac{5}{2} g^6 \zeta_4 + g^8 \left(
     \tfrac{413}{8} \zeta_6 - 2 \zeta_3^2 \right) + O(g^{10}) \,,\\
c_3 &= -\tfrac{35}{8} g^8 \zeta_6 + O(g^{10}) \,,\nonumber
\end{align}
where $\zeta_n = \sum_{k=1}^\infty k^{-n}$ is the Riemann zeta value.

We can derive the decomposition (\ref{eq:R7_origin_line})
to all loop orders via a ``baby'' amplitude boostrap, using
the following conditions:
\begin{enumerate}
\item We assume that $R_7$ is QL.
\item Continuity: The result at $\Or^{(7)}_7$ ($\Or^{(7)}_1$)
is obtained from that on {\sc Line 71} by setting $\lu_7\to0$ ($\lu_1\to0$).
\item Three conditions from dihedral symmetry:
\begin{itemize}
  \item The full $D_7$ is broken on the line but a single reflection
    (flip) survives: $u_i\leftrightarrow u_{8-i}$.
    It exchanges the two end points $u_7=1$ and $u_1=1$.
  \item There is a flip symmetry at $u_7=1$: $u_i\leftrightarrow u_{7-i}$.
  \item The behaviors at the two endpoints are related by cycling
     $u_i\mapsto u_{i+1}$.
\end{itemize}
\item The final-entry (FE) condition.
\end{enumerate}
MHV amplitudes obey a FE condition, which controls their
first derivatives~\cite{Caron-Huot:2011dec}.
For $n=6$ and general kinematics, the FE condition removes three
of the nine symbol letters~\cite{Goncharov:2010jf}, namely $1-u_i$;  but at the origin these letters
are irrelevant because they approach $1$. Hence the six-point FE condition
trivializes at the origin.

In contrast, the seven-point FE condition allows 14 symbol letters for
general kinematics~\cite{Drummond:2014ffa}, which collapse on {\sc Line 71} to six letters out of a
total of seven. We obtain a single constraint,
\be \label{eq:R7_FE_line}
 \bigl[ u_7 \partial_{u_7}
      + u_1 \partial_{u_1} 
      - u_4 \partial_{u_4} \bigr] R_7 = 0\, ,
\ee
where derivatives for $\ln u_7$ are taken independently of $\ln u_1$,
despite the constraint $u_7+u_1=1$ on {\sc Line 71}.
Combining all constraints,
the only allowed QL polynomials are exactly the three
given in (\ref{eq:R7_Pi}), and no linear-logarithmic structures survive.
That is, the possible kinematic dependence of
$R_7$ is already saturated by (\ref{eq:R7_Pi}) at four loops.
We will see that the TBA at strong coupling leads to precisely the
same three $P_i^{(7)}$, and to a natural conjecture
for all higher-loop corrections to the coefficients,
which matches (\ref{eq:R7_cs}) through four loops.

The symbol of the eight-point remainder function $R_8$
is known at two and three loops~\cite{Caron-Huot:2011zgw,Li:2021bwg};
it vanishes at all the origins and interpolating surfaces, as it must to be QL. 
For all the kinematics in table~\ref{tab:octoriginsurfacesrelation},
we computed the full functions at two loops~\footnote{%
We thank A.~McLeod for confirming one of these limits using
results in ref.~\cite{Golden:2021ggj}.}
and, in some cases, up to five loops using the
pentagon operator product expansion (OPE)~\cite{Basso:2013vsa}.
In all cases, we found that the remainder function $R_8$ is QL~\footnote{
We have checked that the full one-loop amplitude, including all
non-dual-conformal terms associated with infrared divergences, is QL
for all $n=6,7,8$ origins.  This suffices to show that the logarithm of
the full amplitude is QL.  The nontrivial statement is that the dilogarithms
of generic arguments cancel out, which requires the use of the five-term
dilog identity for {\sc Pentagons 234 and 345}.}.

Furthermore, we repeated the all-loop seven-point analysis at eight points,
starting on {\sc Cube 6789}, and then going on to other adjacent regions,
using continuity at the boundaries between regions, see figure~\ref{fig:octbdy}.
In all cases, we found precisely five independent QL polynomials obeying
the restrictions.
On {\sc Cube 6789}, see table~\ref{tab:octoriginsurfacesrelation},
they have the form,
\be \label{eq:eight_all_orders}
R_8({\hbox{\sc Cube 6789}}) = \sum_{i=1}^5 d_i P_i^{({\sc C})} \,,
\ee
where,
\begin{align}
P_1^{({\sc C})}
&= \sum_{i=1}^8 \lu_i^2 - 2\sum_{i=1}^4 \lu_i\lu_{i+4} - 2(\lu_3-\lu_7)(\lu_4-\lu_8) \,, \label{eq:cubeP1}\\
P_2^{({\sc C})}
&= 2\sum_{i=1}^4 \lu_i\lu_{i+4} + (\lu_3-\lu_7)(\lu_4-\lu_8) + (\lu_1+\lu_5)(\lv_1+\lv_4) \nonumber\\
&+ (\lu_3+\lu_4+\lu_7+\lu_8)\lv_3 + (\lu_2+\lu_6)\lv_2 + \sum_{i=1}^4 \lv_i^2 \,,
\label{eq:cubeP2}
\end{align}
with $\lu_i \equiv \ln u_i$ and $\lv_i \equiv \ln v_i$.
The lengthier $P^{({\sc C})}_{3,4,5}$ are provided in
appendix~\ref{appx:oct-cube-6789}.
One has $d_{3} = d_{4} = \zeta_{2} g^{4}$ through two loops;
the remaining coefficients start at higher orders.
The same form~\eqref{eq:eight_all_orders} applies in the other
QL-connected regions, with the same $d_{i}$'s but different polynomials.
Similarly, the baby bootstrap yields a five-polynomial ansatz for
{\sc Superline 1}; since it is disconnected from the other regions,
it comes with its own set of coefficients, $f_i$.  We give the expressions for all five polynomials in all possible regions,
along with weak coupling expansions of the $d_i$ and $f_i$ coefficients
through eight loops,
in the ancillary files {\tt octagon\_QL\_formula.txt} and
{\tt octagon\_QL\_coefs.txt}.

\section{Master formula from TBA}
Additional insight into the QL behavior of the amplitudes may be found at
strong coupling using the AdS/CFT-dual string theory description, which
maps the problem to computing the minimal world-sheet area for a string
anchored on a null polygonal contour at the boundary of AdS~\cite{Alday:2007hr}.
Using the integrability of the classical string theory~\cite{Bena:2003wd},
it boils down to solving a set of non-linear TBA
integral equations~\cite{Alday:2009dv,Alday:2010vh}.
We will now outline how the TBA equations can be linearized near origins.
A (weighted) Fourier transformation from the TBA spectral parameter $\theta$ 
to a variable $z$, related to the tilt angle, converts the integral equations
to a simple matrix equation, and allows us to express the minimal area
(the logarithm of the strong-coupling amplitude) as a single integral over $z$.
The crux of our finite-coupling conjecture is to move the 't Hooft coupling
$\sqrt{\lambda}$ inside the integral and absorb it into the tilted cusp
anomalous dimension.  The resulting master formula~(\ref{eq:master})
can be evaluated either at finite coupling, or at weak coupling where it agrees
with all the perturbative data reviewed above.

For the TBA analysis, we use coordinates
$\{\sigma_{s}, \tau_{s}, \varphi_{s}\}$, $s = 1, \ldots, n-5$,
originally developed for analyzing the OPE~\cite{Alday:2010ku,Basso:2013vsa}.
The TBA equations are for a family of $3(n-5)$ functions $Y_{a, s}(\theta)$,
with $a = \{0, \pm 1\}$~\cite{Alday:2010ku,Bonini:2015lfr}:
\beq\label{eq:TBA}
\begin{aligned}
&\ln{Y}_{a, s}(\theta) = I_{a, s}(\theta) \\
&\,\,\, +  \sum_{b, t}\int\limits \frac{k_{a}(\theta)d\theta'}{2\pi k_{b}(\theta')} K_{a, s}^{b, t}(\theta-\theta') \ln{(1+Y_{b, t}(\theta'))}\, ,
\end{aligned}
\eeq
where the sum runs over $b = 0, \pm 1$, $t = s, s\pm 1$, with $k_{a}(\theta) = i^a \sinh{(2\theta - i\pi a/2)}$ and for some kernels $K$.
The driving terms $I_{a, s}$ encode the cross ratios, and are given explicitly
in terms of the OPE coordinates,
\beq
I_{a, s}(\theta) = a\varphi_{s} - m_{a}\tau_{s} \cosh{\theta}
+ (-1)^{s} i m_{a} \sigma_{s} \sinh{\theta}\, ,
\eeq
with $m_{a} = 2\cos{(a\pi/4)}$.
The dependence on the hyperbolic angle $\theta$ corresponds to a
collection of interacting relativistic particles, of mass $m_{a}$ and
charge $a$, coupled to various temperatures  $1/\tau_{s}$ and chemical
potentials $\varphi_{s}$.

Drawing inspiration from the hexagon ($n=6$) analysis~\cite{Basso:2020xts,Ito:2018puu}, we expect origins to map to extreme limits where the particles are
subject to large chemical potentials, $|\varphi_{s}| \rightarrow \infty$,
and to small temperatures, $\tau_{s}\rightarrow \infty$. There are several ways of taking limits for $n>6$. We may send each $\varphi_{s}$ to either $+\infty$ or $-\infty$,
with each case labelled by a sequence
$\Sigma_{n} = (h_{1}, \ldots , h_{n-5})$
with $h_{s} = \varphi_{s}/|\varphi_{s}|$.
In such limits, we expect the particles with $a = h_{s}$ to condense,
and the remaining ones to decouple.
Namely, for a given choice $\Sigma_{n}$, we assume that
$Y_{a, s}(\theta) \gg 1$ if $a = h_s$ and $Y_{a, s} = 0$ otherwise, and linearize eq.~\eqref{eq:TBA} using
$\ln{(1+Y_{b, t})} \rightarrow \delta_{b, h_{t}} \ln{Y_{h_{t}, t}}$. We also assume that the above conditions hold
over the entire real $\theta$ axis.

The problem may then be solved by going to Fourier space.
One defines
\beq
\hat{f}(z) = \int\limits_{-\infty}^{\infty}
\frac{d\theta}{2\pi \cosh{(2\theta)}} z^{2i\theta/\pi} f(\theta)\, ,
\eeq
with a measure introduced to eliminate the weight in eq.~\eqref{eq:TBA}
and with the Fourier variable $(2\ln{z})/\pi$, with $z >0$,
introduced to rationalize all expressions.
Setting $Y_{s} = Y_{h_{s}, s}, I_{s} = I_{h_{s}, s}$, eq.~\eqref{eq:TBA} yields
\beq
\widehat{\ln{Y}_{s}}(z) =
\sum_{t=1}^{n-5}\left[1-K_{n}(z)\right]^{-1}_{s, t}\,  \hat{I}_{t}(z)\, ,
\eeq
with the square matrix $(K_{n}(z))_{s, t} =
\int \frac{d\theta}{2\pi} K_{h_{s}, s}^{h_{t}, t}(\theta) z^{2i\theta/\pi}$.
At strong coupling, $\sqrt{\lambda} = 4\pi g \gg 1$, the remainder function
is given by the TBA free energy~\cite{Alday:2009dv,Alday:2010vh,Alday:2010ku},
which becomes
\bea
R_{n}^{\textrm{string}} &=&
-\frac{\sqrt{\lambda}}{\pi^{2}} \int_{0}^{\infty}
\frac{dz}{z} \mathcal{S}_{n}(z) + \dots\,,  \label{eq:string}\\
\cS_{n}(z) &\equiv& \sum_{s=1}^{n-5}
\widehat{I}_{s}(1/z) \widehat{\ln{Y}_{s}}(z) \,. \label{eq:Sn_def}
\eea
The ellipses stand for a simple term $\propto \Gamma_{\textrm{cusp}}$, to
which we shall return shortly. Importantly, the integrand $\cS_{n}(z)$ is a
rational function of $z$. For any limit $\Sigma_{n}$, it may be cast into
the form (see appendix~\ref{appx:TBA} for details)
\beq\label{eq:poles}
\cS_{n}(z) = \frac{z(1-z^{3})\cP^{\Sigma}_{n}(z)}{(1+z)(1+z^{2})(1-z^{3(n-4)})}\, ,
\eeq
where $\cP^{\Sigma}_{n}(z) = z^{3n-14}\cP^{\Sigma}_{n}(1/z)$ is a polynomial of degree
$3n-14$ in $z$ and is quadratic
in $\{\sigma_{s}, \tau_{s}, \varphi_{s}\}_{s = 1, \ldots , n-5}$.

Eq.~(\ref{eq:string}) may be turned into an all-order
conjecture by bringing $\sqrt{\lambda}$ under the integral sign and
promoting it to a full function of the variable $z$. To be precise,
we conjecture that $R_n$ takes at finite coupling the form of a
contour integral in the dual variable $z$,
\beq\label{eq:master}
R_{n} = -\frac{1}{2} \oint_{C_{n}}
\frac{dz}{2\pi i z} (z-1/z) \tilde{\G}(z, g) \mathcal{S}_{n}(z)\, ,
\eeq
with $\tilde{\G}(z, g) = \G(z, g) -\Gc(g)$ and with $\G(z, g)$
the tilted cusp anomalous dimension, viewed here as a function of
$z = -e^{2i\alpha}$,
\beq \label{eq:Gamma_alpha_def}
\G(z , g) = \Gamma_\alpha(g)\, .
\eeq
Eq.~\eqref{eq:master} neatly factorizes the coupling dependence,
which resides in $\G(z, g)$, and the kinematics, which sits in the string
integrand $\cS_n(z)$. The contour $C_{n}$ is a sum of small circles around
the singularities of $\mathcal{S}_{n}(z)$; from eq.~\eqref{eq:poles} they 
are poles on the unit circle $|z| = 1$, mapping to real angles $\alpha$.
The original string formula is recovered by using the strong coupling
behavior~\cite{Basso:2020xts}
\beq
\Gamma_{\alpha} \approx \frac{2\alpha\sqrt{\lambda}}{\pi^{2}\sin{(2\alpha)}}
\,\,\,\, \Rightarrow \,\,\,\,
\G(z) \approx -\frac{2\sqrt{\lambda}\log{(-z)}}{\pi^2 (z-1/z)}\,.
\eeq
The integral in eq.~\eqref{eq:string} follows from the term
$\propto \G(z)$, by wrapping the contour on the logarithmic cut along $z>0$,
whereas the term $\propto \Gamma_{\textrm{cusp}} = \Gamma_{\pi/4}$
accounts for the ellipses in eq.~\eqref{eq:string}.

At finite coupling, one may calculate eq.~\eqref{eq:master} by residues,
around the poles in eq.~\eqref{eq:poles}, and write
\beq\label{eq:Rn-final}
R_{n} = \sum_{\alpha} \tilde{\Gamma}_{\alpha}(g)
\times P^{\Sigma_{n}}_{\alpha} (\{\sigma_{s}, \tau_{s}, \varphi_{s}\})\, ,
\eeq
with $\tilde{\Gamma}_{\alpha} = \Gamma_{\alpha}-\Gamma_{\textrm{cusp}}$
and with the sum running over
\beq\label{eq:alphas}
\alpha = \frac{\pi}{2}-\frac{\pi p}{3}-\frac{\pi k}{3(n-4)} \, ,
\eeq
with $k = 1, \ldots , n-5$ and $p=0,1,2$.
The associated polynomials $P^{\Sigma_{n}}_{\alpha}$ follow straightforwardly
from the TBA analysis, but are too bulky to be shown here (see eq.~\eqref{eq:polynomial}).
At last, one may eliminate the OPE parameters in favor of the cross ratios,
using general formulae in ref.~\cite{Basso:2013aha}.
In the limit $|\varphi_{s}|\gg \tau_{s} \gg 1$, with $\sigma_{s}$ held fixed,
these relations reduce to simple mappings between the OPE parameters and
the logarithms of the cross ratios.

For illustration, when $n=6$, one finds
\beq\label{eq:use-for-6}
u_{1} \approx e^{\tau+\sigma-|\varphi|}\, , \qquad u_{2}\approx e^{-2\tau}\, ,
\qquad u_{3} \approx e^{\tau-\sigma-|\varphi|}\, ,
\eeq
and eq.~\eqref{eq:Rn-final} and $P^{\Sigma_{6}}_{\alpha}$ give
\beq
R_{6} = -\sum_{\alpha = 0, \pm \pi/3}
\frac{\tilde{\Gamma}_{\alpha}(g)}{24}
|l_{1} + e^{2i\alpha}l_{2} + e^{4i\alpha}l_{3}|^2 \, ,
\eeq
in perfect agreement with ref.~\cite{Basso:2020xts}, using $\tilde{\Gamma}_{-\alpha} = \tilde{\Gamma}_{\alpha}$.
For $n=7$, one gets
\beq\label{eq:use-for-7}
\begin{aligned}
&u_{1} \approx e^{\tau_{2}-\sigma_{2}-|\varphi_{2}|}\, , &u_{2} \approx e^{-2\tau_{2}}\, , &&u_{3}u_{7} \approx e^{\tau_{2}+\sigma_{2}-|\varphi_{2}|} \, ,\\
&u_{6} \approx e^{\tau_{1}-\sigma_{1}-|\varphi_{1}|}\, , &u_{5} \approx e^{-2\tau_{1}}\, , &&u_{4}u_{7} \approx e^{\tau_{1}+\sigma_{1}-|\varphi_{1}|} \, ,
\end{aligned}
\eeq
with $u_{7} = 1$ for $\Sigma_{7} = (+, +)$,
and $u_{3}+u_{4} = 1$ for $\Sigma_{7} = (+, -)$,
corresponding, respectively, to the origin $\Or_7^{(7)}$ and a cyclic image of {\sc Line 71}.
Using $P^{\Sigma_{7}}_{\alpha}$, we find a perfect agreement with the
general decomposition for the heptagon line, eq.~\eqref{eq:R7_origin_line},
with $c_{3} = a_{3}-a_{1}/2, c_{2} = -a_{3}, c_{1} = a_{2}-a_{1}/2$, where
\beq
a_{j} = \frac{(-1)^{j}}{3\sqrt{3}} \sum_{k=1}^{3} (-1)^{k} \sin{(2\alpha_{k})}\cos{(2(j-1)\alpha_{k})} \tilde{\Gamma}_{\alpha_{k}}(g)\, ,
\eeq
and $\alpha_{1,2,3} = \{\pi/18, 5\pi/18, 7\pi/18\}$. The coefficients
agree with the perturbative results~\eqref{eq:R7_cs}, taking into account
the weak-coupling expansion of the tilted cusp anomalous
dimension~\cite{Basso:2020xts},
$\Gamma_{\alpha}(g) = 4g^{2} -16\zeta_{2}g^{4} \cos^{2}{\alpha}+ O(g^{6})$,
as discussed further in appendix~\ref{appx:tilted-cusp}.

One may proceed similarly for $n = 8$ using $\Sigma_{8} = (+,+,+), (+,+,-), (+,-,+)$ and find three domains describing, respectively, the origin ${\rm O}_{9}$, a line ${\rm O_{3}}$--${\rm O_{4}}$, and a square ending on ${\rm O_{8}, O_{9}}$ and two images of ${\rm O_{7}}$. In all of these cases,
we found perfect agreement with the perturbative results,
with the coefficients matching the two-loop predictions and
the five-loop OPE results.

This analysis does not exhaust all the origins and domains given in table~\ref{tab:octoriginsurfacesrelation}. For example, for
(an image of) {\sc Cube 6789} it covers but a single face.
To reach the missing domains, one should look at a broader class of scalings,
where not only $\varphi_{s}$ and $\tau_{s}$ are allowed to be large
but also $\sigma_{s}$.
These scalings are harder to address in general, because the limit $|\sigma_{s}| \rightarrow \infty$ generates large fluctuations in the $Y$ functions, making it hard to decide which of them are large and which are small. It may also trigger new exceptional solutions, with more particle species condensing simultaneously. In appendix~\ref{appx:superline}, we argue that this happens at $n=8$ for {\sc Superline 1}; we conjecture that its QL behavior is captured by a system of linearized TBA equations based on 4 large $Y$ functions.

\section{Conclusions}
In this letter we initiated a systematic
exploration of origins:
kinematical points and interpolating higher-dimensional surfaces
where high-multiplicity MHV scattering amplitudes in planar $\N=4$ SYM
simplify dramatically and can be predicted (conjecturally) at finite coupling.
Cluster algebras provide a roadmap to the kinematics, while the TBA
and the tilted cusp anomalous dimension $\Gamma_\alpha(g)$ both play a
central role in the master formula for the leading singular behavior.
We expect further kinematical richness to emerge for $n>8$,
based on the appearance of the super-origin $\Or_1$ at $n=8$,
which is not connected (by any QL lines) to the other eight-point origins.
We also have not ruled out the possibilities of even more kinematic boundaries
of the positive region with QL behavior, especially for $n\ge8$.
The behavior in all these regions will certainly play a key role in
constraining the all-orders behavior of MHV amplitudes
for generic kinematics.  Our findings may also have implications for
other planar ${\cal N}=4$ observables,
such as correlators of large-charge operators,
which exhibit QL behavior for small cross
ratios~\cite{Coronado:2018cxj,Kostov:2019auq,Belitsky:2019fan,Bercini:2020msp}.
The great similarity between the two problems suggests that a similar origin
story, with a rich pattern of limits and tilted cusp anomalous dimensions,
may be uncovered for all these higher-point functions.


\begin{acknowledgments}
We are grateful to Niklas Henke, Gab Dian, Andrew McLeod, Amit Sever and Pedro Vieira for interesting discussions. The work of BB was supported by the French National Agency for Research grant ANR-17-CE31-0001-02. The work of LD and YL was supported by the US Department of Energy under contract DE--AC02--76SF00515. YL was also supported in part by the Benchmark Stanford Graduate Fellowship, the Heising-Simons Foundation, the Simons Foundation, and National Science Foundation Grant No. NSF PHY-1748958. GP acknowledges support from the Deutsche Forschungsgemeinschaft under Germany's Excellence Strategy – EXC 2121 ``Quantum Universe'' – 390833306. BB, YL and LD thank the Kavli Institute for Theoretical Physics for hospitality. 
\end{acknowledgments}

\onecolumngrid

\begin{figure}[h]
\begin{minipage}{0.5\textwidth}
		\begin{tikzpicture}[scale=1.1]
		\node at (0,3) (b11) {$\frac{\ab{1234}\ab{1256}}{\ab{1236}\ab{1245}}$};
		\node at (0,2) (b21) {$\frac{\ab{1235}\ab{1456}}{\ab{1256}\ab{1345}}$};
		\node at (0,1) (b31) {$\frac{\ab{1245}\ab{3456}}{\ab{1456}\ab{2345}}$};

	\node at (2.2,3)  (b12) {$\frac{\ab{1235}\ab{1267}}{\ab{1237}\ab{1256}}$};
		\node at (2.2,2)  (b22) {$\frac{\ab{1236}\ab{1245}\ab{1567}}{\ab{1235}\ab{1267}\ab{1456}}$};
		\node at (2.2,1)  (b32) {$\frac{\ab{1256}\ab{1345}\ab{4567}}{\ab{1245}\ab{1567}\ab{3456}}$};

		\node at (4,3)  (b13) {\ldots};
		\node at (4,2)  (b23) {\ldots};
		\node at (4,1)  (b33) {\ldots};

		\node at (7.2,3)  (b14) {$\frac{\ab{123\,n-2}\ab{12\,n-1\,n}}{\ab{123\,n}\ab{12\,n-2\,n-1}}$};
		\node at (7.2,2)  (b24) {$\frac{\ab{123\,n-1}\ab{12\,n-3\,n-2}\ab{1\,n-2\,n-1\,n}}{\ab{123\,n-2}\ab{12\,n-1\,n}\ab{1\,n-3\,n-2\,n-1}}$};
		\node at (7.2,1)  (b34) {$\frac{\ab{12\,n-2\,n-1}\ab{1\,n-4\,n-3\,n-2}\ab{n-3\,n-2\,n-1\,n}}{\ab{12\,n-3\,n-2}\ab{1\,n-2\,n-1\,n}\ab{n-4\,n-3\,n-2\,n-1}}$};
		\draw[->](b11)--(b21);
		\draw[->](b21)--(b31);
		\draw[->](b11)--(b12);
	    \draw[->](b21)--(b22);
	    	\draw[->](b31)--(b32);
        	\draw[->](b22)--(b11);
		\draw[->](b32)--(b21);
		\draw[->](b12)--(b22);
		\draw[->](b22)--(b32);
		\draw[->](b21)--(b31);
		\draw[->](b12)--(b13);
	    \draw[->](b22)--(b23);
	    	\draw[->](b32)--(b33);
        	\draw[->](b23)--(b12);
		\draw[->](b33)--(b22);
		
		\draw[->](b14)--(b24);
		\draw[->](b24)--(b34);
		\draw[->](b13)--(b14);
	    \draw[->](b23)--(b24);
	    	\draw[->](b33)--(b34);
		\draw[->](b24)--(b13);
		\draw[->](b34)--(b23);
	\end{tikzpicture}
\end{minipage}
\begin{minipage}{0.49\textwidth}
\hfill \includegraphics[width=0.72\textwidth]{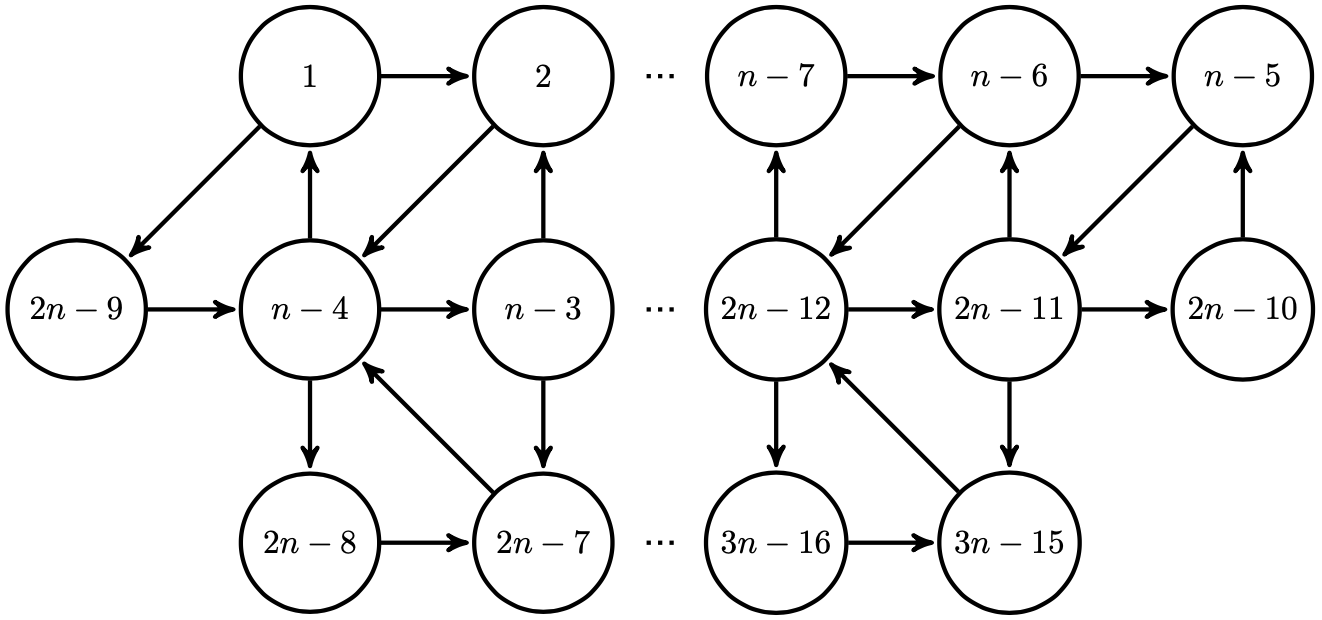}
\end{minipage}
\caption{Left: Initial cluster of the ${\rm Gr}(4,n)$ cluster algebra, with respect to  $\mathcal{X}$-coordinates. Right: Choice of cluster we begin mutating from so as to obtain all contiguous origin limits. The numbering of vertices is inherited from the initial cluster, where it starts at the top left, and increases first as we change columns, and then as we change rows.}
\label{fig:Gr4nClusterX}
\end{figure}
\twocolumngrid

\appendix

\section{Tilted Cusp Anomalous Dimension}\label{appx:tilted-cusp}

The \textit{tilted} cusp anomalous dimension was introduced in
ref.~\cite{Basso:2020xts} using a one-parameter deformation of the
BES equation~\cite{Beisert:2006ez}:
\beq \label{eq:Gamma_alpha}
\Gamma_\alpha(g) = \Gamma(\alpha, g)
= 4g^2 \left[1+\K(\alpha, g)\right]^{-1}_{11}\, ,
\eeq
where $\K(\alpha, g)$ is a semi-infinite matrix~\cite{Benna:2006nd}
with elements
\beq
\begin{aligned}
\K(\alpha, g)_{ij}
&=  4j(-1)^{ij+j}\cos{\alpha} \int\limits_{0}^{\infty}\frac{dt}{t}
  \frac{J_{i}(2gt)J_{j}(2gt)}{e^{t}-1} \\
   & \qquad \times \begin{cases}
      \cos{\alpha}\, , & \text{for}\ |i-j| \ \textrm{even} \\
      \sin{\alpha}\, , & \text{otherwise}
    \end{cases}\, , 
\end{aligned}
\eeq
with $i, j \in \mathbb{N}$, and
$J_{i}$ is the $i$-th Bessel function of the first kind. The `11' subscript in eq.~\eqref{eq:Gamma_alpha} refers to the 1,1 entry of the inverse of the matrix.

The deformation parameter $\alpha$ only enters the kernel in the cosine
prefactors.  The usual, undeformed BES equation is recovered when
$\alpha = \pi/4$, $\Gc (g) = \Gamma(\pi/4, g)$.
The first few terms in the weak coupling expansion of $\Gamma_{\alpha}(g)$ are
\vspace{-5pt}
\bea
\Gamma_{\alpha}(g) &=& 4g^{2} \Bigl\{ 1 - 4 \zeta_2 c^2 g^2
+ 8 c^2 (3+5 c^2) \zeta_4 g^4  \nonumber\\
&&\hskip-0.4cm\null
- 8 c^2 \Bigl[ ( 25+42 c^2+35 c^4 ) \zeta_6 + 4 s^2 (\zeta_3)^2 \Bigr] g^6
\nonumber\\
&&\hskip-0.4cm\null
+ 8 c^2 \Bigl[ \Bigl( 245 + \tfrac{1273}{3} c^2 + 420 c^4
                          + \tfrac{700}{3} c^6 \Bigr) \zeta_8  \nonumber\\
&&\hskip-0.1cm\null
            + 16 \zeta_3 ( 5 \zeta_5 + 2 \zeta_2 \zeta_3 c^2 ) s^2 \Bigr] g^8
            + {\cal O}(g^{10}) \Bigr\} \,,~~~~
\label{eq:G_alpha_weak}
\eea
where $c = \cos\alpha$, $s=\sin\alpha$.
We provide the values of $\Gamma_{\alpha}(g)$ through eight loops
in the ancillary file {\tt Gamma\_alpha.txt}.

We remark that, despite the generically irrational trigonometric factors
of $\cos\alpha$ and $\sin\alpha$ in the weak coupling expansion
of an individual $\Gamma_{\alpha}$, in the full sum over angles,
given by eq.~\eqref{eq:Rn-final}, there are trigonometric
identities that result in only rational coefficients multiplying the
zeta values in the $c_i$, $d_i$ and $f_i$.  The rationality of the
coefficients can be made manifest to any loop order by an alternate
evaluation of the contour integral in eq.~\eqref{eq:master}, as we now explain.
To any order in perturbation theory, $\G(z,g)$ has poles only at
$z=0,\infty$.  Therefore, the contour $C_{n}$ can be deformed away from
the unit-circle poles of $\cS_{n}(z)$,
so that it encircles $z=0$ and $z=\infty$ instead.  A symmetry under
$z \leftrightarrow 1/z$ ensures that the $z=\infty$ residue equals
the one at $z=0$, resulting in
\beq
R_{n} \overset{\textrm{PT}}{=} \oint \frac{dz}{2\pi i z}(z-1/z)\tilde{\G}(z, g) \cS_{n}(z)\, ,
\eeq
with the contour going about $z=0$.
From the perturbative expansion of $\G(z, g)$,
which follows from eq.~\eqref{eq:G_alpha_weak} by letting
$c^2 = -\tfrac{1}{4} (z+z^{-1}-2)$,
$s^2 =  \tfrac{1}{4} (z+z^{-1}+2)$,
it is clear that only rational coefficents will appear in the
residue at $z=0$.  The $z=0$ residue evaluation is also the simplest
way to compare the master formula with perturbative data.

\section{Gram Determinant Constraints}\label{appx:Gramdet}

At seven points, the seven cross ratios $u_i \equiv u_{i+1,i+4}$
(with all indices mod 7)
obey a single Gram determinant constraint~\cite{Dixon:2020cnr},
\begin{widetext}
\bea\label{eq:gram7}
0 &=& 1 + \biggl[-u_1 + u_1 u_3 + u_1 u_4 + u_1 u_2 u_5 - u_1 u_3 u_5
- u_1^2 u_4 u_5 - 2 u_1 u_2 u_4 u_5
+ u_1 u_2 u_3 u_5 u_6 + u_1^2 u_2 u_4 u_5^2
\ +\ \text{cyclic}\biggr]\nonumber\\
&&+ u_1 u_2 u_3 u_4 u_5 u_6 u_7 \,.
\eea
\end{widetext}
At eight points, there are 12 cross ratios, eight $u_i \equiv u_{i+1,i+4}$
and four $v_i \equiv u_{i+1,i+5}$.  They obey three independent
Gram determinant constraints, which are provided in the ancillary file {\tt octagon\_Gram.txt}.

\section{Cluster Origins}\label{appx:kinematics}

After briefly reviewing the positive region and its cluster algebra structure, in this appendix we provide further details on how the latter can be used in order to classify origin limits.

The space of dual conformal $n$-particle kinematics of $\N=4$ SYM amplitudes is most conveniently described in terms of $n$ cyclically ordered \emph{momentum twistors} $Z_i\in \mathbb{CP}^3$~\cite{Hodges:2009hk}, which can be assembled in a $4\times n$ matrix. The conventional Mandelstam invariants of eq.~\eqref{eq:u_def}, for example, may be expressed in terms of certain maximal minors of this matrix,
\begin{equation}\label{xToZ}
x^2_{ij} \propto \langle i\!-\!1\, i\, j\!-\!1\, j \rangle\,,
\end{equation}
\be\label{fourbrak}
\ab{ijkl} \equiv\langle Z_{i} Z_j Z_{k} Z_l \rangle=\det(Z_i Z_j Z_k Z_l)\, ,
\ee
up to proportionality factors that drop out from conformally invariant quantities. The \emph{positive region} of this space~\cite{Arkani-Hamed:2012zlh,Golden:2013xva}, which closely resembles the ${\rm Gr}(4,n)$ Grassmannian, is defined as the subspace where
\be
\ab{ijkl}>0\qquad \forall\,\, i<j<k<l\,,
\ee
and it is naturally endowed with a cluster algebra structure, as is reviewed for example in ref.~\cite{Papathanasiou:2022lan}. 

The building blocks of cluster algebras are \emph{cluster variables}, which are grouped into overlapping subsets (the \emph{clusters}) of the same size (the \emph{rank} of the cluster algebra). Starting from an initial cluster, cluster algebras may be constructed recursively by a \emph{mutation} operation on the cluster variables.

The cluster variable content and mutation rule of each cluster may be encoded in the vertices of a quiver, and the arrows connecting them, respectively.  The initial quiver of the ${\rm Gr}(4,n)$ cluster algebra is depicted at the left of figure~\ref{fig:Gr4nClusterX}, where it is evident that the rank coincides with the dimension of the kinematic space, $3n-15$. While the original definition of cluster algebras by Fomin and Zelevinsky is with respect to so-called cluster $\A$-coordinates~\cite{1021.16017,1054.17024}, for the purposes of this paper we will be exclusively using the closely related cluster $\mathcal{X}$-coordinates introduced by Fock and Goncharov~\cite{FG03b}. The reason is that for each cluster, these variables $\mathcal{X}_i$ correspond to the coordinates of a chart describing a compactification of the positive region (whose interior maps to $0<\mathcal{X}_i<\infty$). They are thus ideally suited for locating origin limits at its boundary.

The arrows between vertices $i$ and $j$ of the quiver define an antisymmetric \emph{exchange matrix} $B$ with components
\be\label{eq:ExchMatrix}
b_{ij}=(\#\ \text{arrows}\,\, i\to j)-(\#\ \text{arrows}\,\, j\to i)\,.
\ee
Upon mutation of the $k$-th vertex of the quiver, the $\mathcal{X}$-coordinates transform as
\be
\label{eq:xMutation}
\mathcal{X}_i' = \begin{cases}
    1/\mathcal{X}_i              &   k = i\,,\\
    \mathcal{X}_i\bigl(1+\mathcal{X}_k^{{-\rm sgn}(b_{ki})}\bigr)^{-b_{ki}} & k \neq i\,,
  \end{cases}
\ee
whereas the components of the exchange matrix in the new cluster, $B'$, are given by
\begin{align}\label{eq:Bmutation}
	b'_{ij} = 
		\begin{cases}
			-b_{ij} \quad &\text{for}\,i=k\,\text{or}\,j=k\\
			b_{ij} + \left[-b_{ik}\right]_+b_{kj} + b_{ik}\left[b_{kj}\right]_+\quad &\text{otherwise}
		\end{cases}\,,
\end{align}
where $\left[x\right]_+ = \max\left(0,x\right)$.

A (`web'-)parametrization of the momentum twistor matrix in terms of the $\mathcal{X}$-coordinates of the initial cluster can be constructed algorithmically for any $n$~\cite{Speyer2005}, see also~\cite{Drummond:2019cxm,Henke:2021ity} for a simplified reformulation, and by virtue of the mutation rule~\eqref{eq:xMutation} also for any other cluster. With the help of eqs.~\eqref{eq:u_def} and~\eqref{xToZ}--\eqref{fourbrak} we may then express all cross ratios in terms of them, and evaluate them at the vertex of the boundary polytope corresponding to each cluster, i.e.~we let all its $\mathcal{X}$-coordinates $\mathcal{X}_i\to 0$. To illustrate this process with a particular example, the web-parametrization of the $4 \times 6$ matrix of momentum twistors of the six-particle amplitude is
\be\left(
\begin{array}{cccccc}
 1 & 0 & 0 & 0 & -1 & -1-\mathcal{X}_1-\mathcal{X}_1 \mathcal{X}_2-\mathcal{X}_1 \mathcal{X}_2 \mathcal{X}_3 \\
 0 & 1 & 0 & 0 & 1 & 1+\mathcal{X}_1+\mathcal{X}_1 \mathcal{X}_2 \\
 0 & 0 & 1 & 0 & -1 & -1-\mathcal{X}_1 \\
 0 & 0 & 0 & 1 & 1 & 1 \\
\end{array}
\right)\,,
\ee
such that the cross ratios may be expressed in terms of the 
$\mathcal{X}$-coordinates of the initial cluster as
\be
\begin{aligned}
u_1&=\frac{1}{1+\mathcal{X}_2+\mathcal{X}_2 \mathcal{X}_3}\,,\quad u_3=\frac{\mathcal{X}_1 \mathcal{X}_2}{1+\mathcal{X}_1+\mathcal{X}_1 \mathcal{X}_2}\,,\\
u_2&=\frac{\mathcal{X}_2 \mathcal{X}_3}{\left(1+\mathcal{X}_1+\mathcal{X}_1 \mathcal{X}_2\right) \left(1+\mathcal{X}_2+\mathcal{X}_2 \mathcal{X}_3\right)}\,.
\label{neq6_100}
\end{aligned}
\ee

As is evident in this $n=6$ example, the initial cluster of figure~\ref{fig:Gr4nClusterX} does not yield an origin point limit when $\mathcal{X}_i\to 0$. (We have defined these limits to have at least $3(n-5)$ cross ratios approaching zero, whereas this cluster is a corner of a multi-soft limit~\cite{DelDuca:2016lad} with only $2(n-5)$ vanishing cross ratios.) Nevertheless, it is easy to show that $(n-5)!$ different origin limits may be obtained from it by mutating all the $\mathcal{X}$-coordinates of its middle row in all possible orders. In other words, these particular origins are $(n-5)$ mutations away from the initial cluster. This simple pattern has been inferred from the $n=6,7$ cases, and additionally checked up to $n=10$. A particularly simple choice of ordering is from left to right, which leads to the quiver at the right of figure~\ref{fig:Gr4nClusterX} for any $n$.

\begin{figure}[t]
\begin{center}
\includegraphics[width=3.55in]{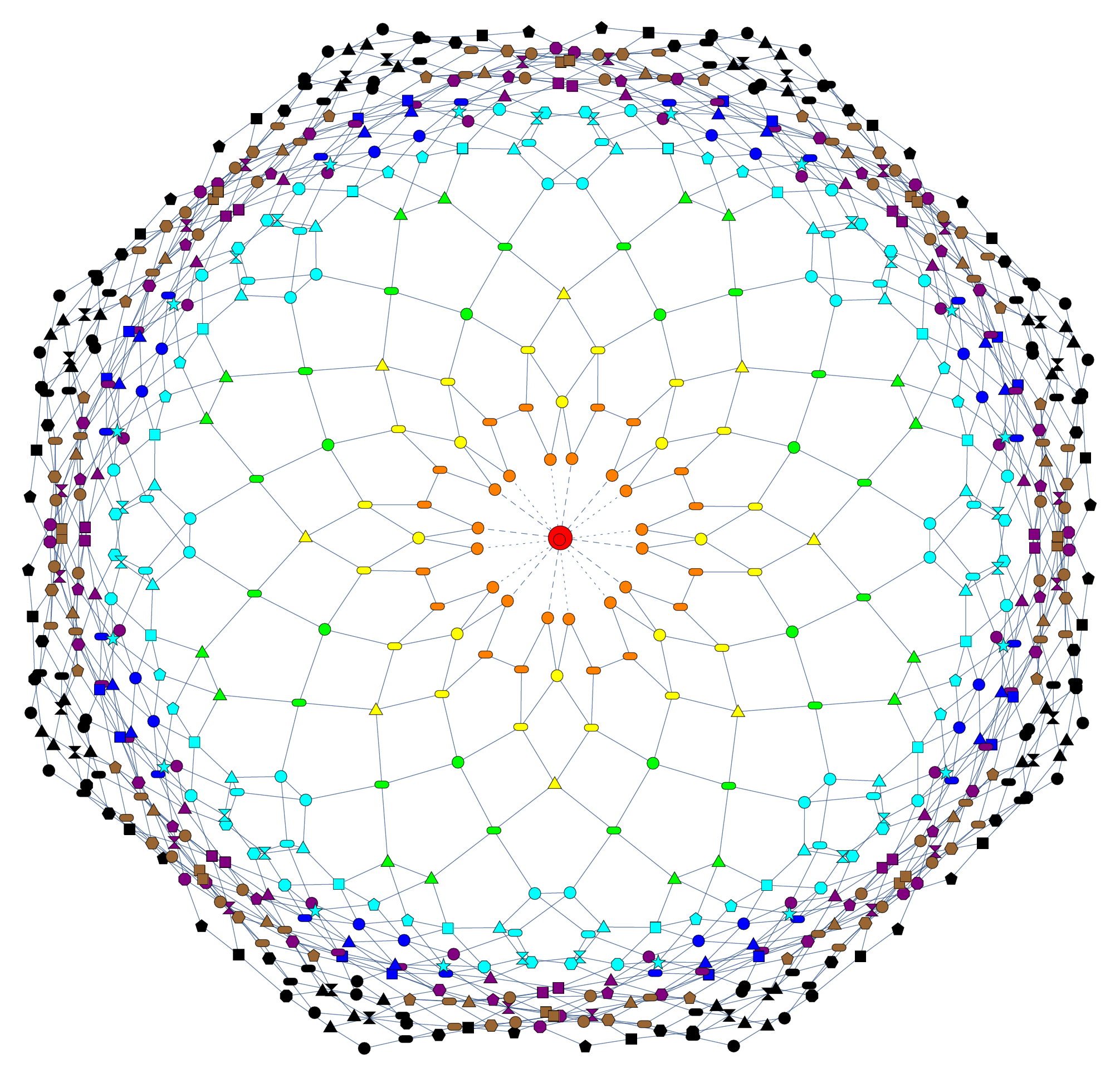} 
\end{center}
\caption{Web of octagon origin clusters, color coded according to the origin classes \textcolor{red}{1},  \textcolor{orange}{2}, \textcolor{Goldenrod}{3}, \textcolor{ForestGreen}{4}, \textcolor{cyan}{5}, \textcolor{blue}{6}, \textcolor{purple}{7}, \textcolor{darkbrown}{8}, 9 in table~\ref{tab:octorigins}. It may be viewed as a half-sphere with two $\textcolor{red}{\rm O_1}$ at the north pole and with $\rm O_9$'s at the equator. The missing half-sphere is the parity image, which has been omitted for simplicity. Lines correspond to mutations between clusters, and same-colored vertices of different shape denote different directions of approach within each origin class. The lines between $\rm \textcolor{red}{O_1}$ and $\rm \textcolor{orange}{O_2}$ are not solid to indicate that the
remainder function is not QL on them.  They are dashed or dotted to
distinguish which of the two overlapping $\rm \textcolor{red}{O_1}$ vertices they start from.
The (super)line between the two $\rm \textcolor{red}{O_1}$'s is not visible.}
\label{fig:O8ClustersParityHalf}
\end{figure}

Starting from this cluster, by further mutating we obtain all other contiguous clusters also corresponding to origin limits, as described in the main text. The \emph{exchange graph} of a cluster algebra is a graph where its clusters are represented by vertices, and the mutations among them by edges. Restricting ourselves just to origin limit clusters and the mutations among them, this partial exchange graph for $n=8$ is depicted in figure~\ref{fig:O8ClustersParityHalf}. The edges of mutations between clusters of the same origin class sometimes amount to a 
change of a cross ratio by finite amount from 0 to 1 or vice versa, and sometimes by an infinitesimal amount. Namely they may connect different dihedral images among the same class, or two different directions of approach to the same strict limit. When approaching origin limits from the interior of the positive region, such that the amplitudes only exhibit QL behavior with respect to the cross ratios as described in the main text, the latter lines play no role, because they become points in the relevant cross ratio space.
(Note that ``direction of approach'' is related to ``speed of approach'', the behavior can depend on the Riemann sheet, and here we are on a Euclidean sheet. For example, on a physical scattering sheet, the limit $\mathcal{X}_i \to 0$ in eq.~\eqref{neq6_100} corresponds to the multi-Regge limit, where the remainder function is definitely {\it not} QL, although it is QL there on the Euclidean sheet.)
We may therefore coarsen the exchange graph by identifying clusters connected by such mutations. Further omitting dihedrally related vertices such that the higher-dimensional limits of table~\ref{tab:octoriginsurfacesrelation} appear only once, we finally arrive at the simplified graph of figure~\ref{fig:octbdy}.

The complete data for the 1188 clusters of ${\rm Gr}(4,8)$ corresponding to origin limits of the eight-particle amplitude, including their exchange matrix and $\mathcal{X}$-coordinates, momentum twistor parametrization and values of the cross ratios in terms of these coordinates, as well as the adjacency matrix recording the mutation connectivity shown in figure~\ref{fig:O8ClustersParityHalf}, may be found in the attached ancillary file \texttt{OctOriginClusterData.m}.

\section{Octagon {\sc Cube 6789}}\label{appx:oct-cube-6789}

On {\sc Cube 6789}, five independent QL polynomials
are allowed by continuity, dihedral symmetry, and FE conditions.
Two of them are given in eqs.~(\ref{eq:cubeP1}) and (\ref{eq:cubeP2}).
The other three are lengthier and are given here:
\begin{align}
P_3^{({\sc C})}
&= \sum_{i=1,2,4,5,6,8} \lu_i\lu_{i+1} - \sum_{i=2,3,6,7} \lu_i\lu_{i+2} \nonumber\\
&+ (\lu_2-\lu_3+\lu_6-\lu_7)\lv_1 + (\lu_3+\lu_7)\lv_2 + (\lu_4+\lu_8)\lv_4 \nonumber\\
&- (\lv_1-\lv_4)(\lv_3-\lv_2) + \lv_2\lv_3 \,,\nonumber\\
P_4^{({\sc C})}
&= \sum_{i=1}^8 \lu_i\lu_{i+2} + \sum_{i=1}^4 (\lu_{i+2}+\lu_{i+3}+\lu_{i+6}+\lu_{i+7})\lv_i \nonumber\\
&- (\lu_2+\lu_6)\lv_1 - (\lu_3+\lu_7)\lv_2 - (\lu_4+\lu_8)\lv_4 \nonumber\\
&+ 2(\lv_1\lv_3 + \lv_2\lv_4) \,,\nonumber\\
P_5^{({\sc C})}
&= \sum_{i=1,2,3,5,6,7} \lu_i\lu_{i+3} + \sum_{i=2,3,6,7} \lu_i\lu_{i+2} \nonumber\\
&+ \sum_{i=1}^4 (\lu_{i-1}+\lu_{i+3})\lv_i
+ \sum_{i=2}^4 (\lu_{i+2}+\lu_{i+6})\lv_i \nonumber\\
&+ \lv_1\lv_3 + \lv_2\lv_4 + \lv_1\lv_2 + \lv_2\lv_3 + \lv_3\lv_4 \,.
\end{align}
We also give the coefficients $d_i$ appearing
in eq.~(\ref{eq:eight_all_orders}) through four loops,
\begin{align}
d_1 &= d_2 = -\tfrac{5}{2} g^6 \zeta_4 + g^8 \left(
    \tfrac{413}{8} - 2 \zeta_3^2 \zeta_6 \right) + O(g^{10}) \,,\nonumber\\
d_3 &= g^4 \zeta_2 - \tfrac{37}{2} g^6 \zeta_4 + g^8 \left(
    \tfrac{1975}{8} \zeta_6 - 2 \zeta_3^2 \right) + O(g^{10}) \,,\nonumber\\
d_4 &= d_3 - \tfrac{35}{8} g^8 \zeta_6 + O(g^{10}) \,,\nonumber\\
d_5 &= O(g^{10}) \,.
\end{align}
We give the values of the $d_i$ through eight loops in the ancillary file {\tt octagon\_QL\_coefs.txt}.

\section{TBA Analysis}\label{appx:TBA}

To construct the remainder function at the $n$-point origins, we need the TBA equations for the charged particles that trigger the QL behavior. Defining
\beq
\begin{aligned}
&f_{\pm, s}(z) = \int\frac{d\theta}{2\pi \cosh{(2\theta)}} z^{2i\theta/\pi} \ln{Y_{\pm 1, s}(\theta)}\, , \\ 
&g_{\pm, s}(z) = \int\frac{d\theta}{2\pi \cosh{(2\theta)}} z^{2i\theta/\pi} \ln{\left[1+Y_{\pm 1, s}(\theta)\right]}\, ,
\end{aligned}
\eeq
and omitting the charge zero particles, eq.~\eqref{eq:TBA} becomes, in Fourier space,
\beq\label{eq:TBA-z}
\begin{aligned}
f_{\pm, s} = \hat{I}_{\pm, s} &+ \frac{z}{1+z^{2}}(g_{+, s}+g_{-, s}) \\
&+ \frac{z^{3}}{(1+z)(1+z^{2})}(g_{\pm, s+1} + g_{\pm, s-1}) \\
&- \frac{z}{(1+z)(1+z^{2})} (g_{\mp, s+1} + g_{\mp, s-1})\, ,
\end{aligned}
\eeq
for $s$ odd. The same equations hold for $s$ even,
after replacing $z\rightarrow 1/z$ in each $z$-dependent coefficient.
The driving terms are known functions of the OPE parameters, given by
\beq
\hat{I}_{\pm, s} =  \frac{\sqrt{z} e_{\pm, s}(z)}{2(1+z)(1+z^2)}\, ,
\eeq
with, for $s$ odd,
\beq
e_{\pm, s} = (\pm \varphi_{s} -\tau_{s} -\sigma_{s}) - 2\tau_{s} z + (\pm \varphi_{s}-\tau_{s}+\sigma_{s}) z^2\, , 
\eeq
and similarly for $s$ even, with $\sigma_{s}\rightarrow -\sigma_{s}$. 

When the chemical potentials and inverse temperatures are large, $|\varphi_{s}|, \tau_{s} \gg 1$, we set $g_{a, s} \rightarrow f_{a, s}$ in eq.~\eqref{eq:TBA-z} if the particles $(a, s)$ condense, and $g_{a, s} \rightarrow 0$ otherwise. The equations are then linear in $f$'s and are controlled by a matrix whose $z$-dependent coefficients may be read off from eq.~\eqref{eq:TBA-z}. To be more concrete, for the choice $\Sigma_{n} = (h_{1}, \ldots , h_{n-5})$, with $h_{s}$ the charge of the condensed particles in the $s$-th OPE channel, the TBA kernels can be packed into a tridiagonal $(n-5)\times (n-5)$ matrix,
\beq\label{eq:tridiagonal}
1- K_{n} = \frac{1}{(1+z)(1+z^2)}\left[\begin{array}{cccc} a_{1} & b_{1} & &  \\ c_{1} & \ddots & \ddots &  \\  & \ddots & \ddots & b_{n-6} \\ & & c_{n-6} & a_{n-5} \end{array}\right]\, ,
\eeq
with  $a_{s}(z) = 1+z^{3}$, $c_{s}(z) = z^{3} b_{s}(1/z)$, for $s \in \{1, \ldots , n-5\}$, and $b_{s} = \frac{1}{2}(1-h_{s}h_{s+1}) z -\frac{1}{2}(1+h_{s}h_{s+1}) z^{3}$, for $s$ odd, and similarly for $s$ even with $b_{s} \rightarrow c_{s}$.

The string integrand $\cS_{n}(z)$ is a quadratic form in the OPE parameters, defined by contracting the inverse of $1-K_{n}(z)$ with the TBA sources,
\beq
\cS_{n}(z) =\hat{I}(1/z) \cdot \left[1-K_{n}(z)\right]^{-1} \cdot \hat{I}(z)^{T}\, ,
\eeq
with $\hat{I}(z) \equiv (\hat{I}_{h_{1}, 1}(z), \ldots , \hat{I}_{h_{n-5}, n-5}(z))$ and $T$ the transpose.
Straightforward algebra with the matrix~\eqref{eq:tridiagonal} allows us to cast $\cS_{n}(z)$ into the canonical form~\eqref{eq:poles} with $\cP_{n}^{\Sigma}(z)$ encoding all the kinematic dependence. One may achieve further simplifications by factorizing $\cP_{n}^{\Sigma}(z)$ on the support of the poles of $\cS_{n}(z)$. Namely, one may show that
\beq\label{eq:S-canonical}
\cP^{\Sigma}_{n}(z) \approx  -\frac{1}{8z}(1+z^{3(n-4)}) Q_{n}^{\Sigma}(z)Q_{n}^{\Sigma}(1/z) \, ,
\eeq
up to terms integrating to zero in the contour integral~\eqref{eq:master}, with
\beq
Q_{n}^{\Sigma}(z) = \sum_{s=1}^{n-5} (-1)^{s+1}  \frac{1-z^{3(n-4-s)}}{1-z^{3}}e_{s}(z) \prod_{i=1}^{s-1}b_{i}(z)\, ,
\eeq
and $e_{s} = e_{h_{s}, s}$. The above function defines a polynomial in $z$, with coefficients depending linearly on $\{\sigma_{s}, \tau_{s}, \varphi_{s}\}_{s = 1, \ldots , n-5}$. (The rhs of eq.~\eqref{eq:S-canonical} is a Laurent polynomial in $z$, unlike $\cP_{n}^{\Sigma}(z)$ which is polynomial in $z$. Both polynomials obey $\cP^{\Sigma}_{n}(z) = z^{3n-14}\cP^{\Sigma}_{n}(1/z)$, which ensures the symmetry under $z\rightarrow 1/z$ of the string integrand~\eqref{eq:poles}.)

For illustration, when $n=6$,
\be
\begin{aligned}
\cS_{6}(z) = \frac{\hat{I}(1/z)\hat{I}(z)}{1-K_{6}(z)} 
 = \frac{z \cP_{6}(z)}{(1+z)(1+z^{2})(1+z^{3})}\, ,
\end{aligned}
\ee
with
\beq
\begin{aligned}
\cP_{6}(z) &= \frac{z^2}{4} e_1(1/z)e_1(z) \\
& = -\frac{(1+z^{6})}{8z} e_1(1/z)e_1(z) + \frac{(1+z^{3})^{2}}{8z}e_1(1/z)e_1(z)\, .
\end{aligned}
\eeq
and $e_1(z) = -l_{1} + z l_{2} -z^{2} l_{3}$.
We may discard the second term because it vanishes on the relevant poles,
when $1+z^{3} = 0$. (Poles at $z\pm i$ are cancelled by the
vanishing of $\tilde{\Gamma}_{\pi/4}$.)

The polynomial $P^{\Sigma_{n}}_{\alpha}$ in $R_{n} = \sum_{\alpha}\tilde{\Gamma}_{\alpha}P_{\alpha}^{\Sigma_{n}}$ (eq.~\eqref{eq:Rn-final})
follows straightforwardly by evaluating the contour integral~\eqref{eq:master} around the unit-circle poles of the string integrand. Using eqs.~\eqref{eq:poles} and \eqref{eq:S-canonical}, one finds
\beq\label{eq:polynomial}
P^{\Sigma_{n}}_{\alpha} =  -\frac{\cos{\alpha}\cos{(3\alpha)}}{12(n-4)\cos{(2\alpha)}}|Q^{\Sigma}_{n}(-e^{2i\alpha})|^2\, ,
\eeq
with $\alpha$ as in eq.~\eqref{eq:alphas}. Simplifying further the expressions for $n=6,7$, using trigonometric identities, eqs.~\eqref{eq:use-for-6} and~\eqref{eq:use-for-7}, one finds
\beq
P^{n=6}_{\alpha} = -\frac{1}{24} \bigg|\sum_{j=1}^{3}e^{2ij\alpha}l_{j}\bigg|^2\, ,
\eeq
with $\alpha = \{0, \pm \pi/3\}$, for the hexagon origin~\eqref{eq:O6_def}, and
\beq
P^{n=7}_{\alpha} = -\frac{\cos{\alpha} \cos{(3\alpha)}}{36\cos{(2\alpha)}} \bigg|\sum_{j=1}^{7}e^{2ij\alpha}l_{j}\bigg|^2\, ,
\eeq
with $\alpha = \{\pm \pi/18, \pm 5\pi/18, \pm 7\pi/18\}$, for {\sc Line 71}~\eqref{eq:Line71_def}. One verifies the agreement with the general formulae reported earlier for these two cases.

Note that both the polynomial~\eqref{eq:polynomial} and the tilted cusp anomalous dimension are symmetric under $\alpha \rightarrow -\alpha$. Hence, one may restrict the sum over angles to $0\leqslant \alpha\leqslant \pi/2$, including a factor of $2$ for all $\alpha \neq \{0, \pi/2\}$.

Notice also that $P_{\alpha}^{\Sigma_{n}}$ has a pole at $\alpha = \pi/4$, due to the cosine in the numerator in eq.~\eqref{eq:polynomial}. Still, the limit $\alpha \rightarrow \pi/4$ is well-defined at the level of the remainder function, since $\tilde{\Gamma}_{\alpha} = \Gamma_{\alpha} - \Gamma_{\pi/4}$ vanishes at this point. Namely,
\beq
\lim_{\alpha\rightarrow \pi/4} \frac{ \tilde{\Gamma}_{\alpha}\cos{\alpha}\cos{(3\alpha)}}{\cos{(2\alpha)}} = \frac{\Gamma'_{\pi/4}}{4}\, ,
\eeq
where $\Gamma'_{\pi/4} \equiv \partial_{\alpha} \Gamma_{\alpha} |_{\alpha = \pi/4} = 16\zeta_{2} g^{4}-256 \zeta_{4} g^{6}+\ldots\,$.
This limiting procedure is relevant whenever $n$ is a multiple of $4$, as one can see from the general expressions for the roots $\alpha$, see eq.~\eqref{eq:alphas}. In particular, $\Gamma'_{\pi/4}$ enters the description of the QL behavior of the octagon amplitude.

To get rid of the OPE parameters, one needs their mapping to the cross ratios in the limits of interest. The results for $n = 6$ and $7$ are given in eqs.~\eqref{eq:use-for-6} and~\eqref{eq:use-for-7}. Here we provide the missing information for $n = 8$. One finds, when $|\varphi_{s}| \gg \tau_{s} \gg 1$, with $\sigma_{s}$ fixed,
\beq
\begin{aligned}
&u_{3} \approx  e^{\tau_{3}-\sigma_{3}-|\varphi_{3}|}\, , \,\,\, u_{2} \approx e^{-2\tau_{3}}\, , \,\,\, u_{1}u_{4}v_{4} \approx e^{\tau_{3}+\sigma_{3}-|\varphi_{3}|}\, , \\
&v_{1} \approx e^{\tau_{2}-\sigma_{2}-|\varphi_{2}|}\, ,  \,\,\,\, v_{2} \approx e^{-2\tau_{2}}\, , \,\,\, u_{4}u_{8}v_{3}v_{4} \approx e^{\tau_{2}+\sigma_{2}-|\varphi_{2}|}\, , \\
&u_{7} \approx e^{\tau_{1}-\sigma_{1}-|\varphi_{1}|}\, ,  \,\,\, u_{6} \approx e^{-2\tau_{1}}\, , \,\,\, u_{5}u_{8}v_{4} \approx e^{\tau_{1}+\sigma_{1}-|\varphi_{1}|}\, ,
\end{aligned}
\eeq
with $u_{4} = u_{8} = v_{4} = 1$ for $\Sigma_{8} = (+,+,+)$, $u_{8} = v_{4} = u_{1}+v_{3} = 1$ for $\Sigma_{8} = (+,+,-)$, and $v_{3}  = u_{4}+u_{5} = u_{1}+u_{8} =1$ for $\Sigma_{8} = (+,-,+)$. As alluded to before, these relations correspond, respectively, to the origin $\rm O_{9}$, a line between (images of) $\rm O_{3}$ and $\rm O_{4}$, and a square connecting (images of) $\rm O_{9}, \rm O_{8}$ and two $\rm O_{7}$.

We may then compare the TBA prediction for $\Sigma_{8} = (+, +, +)$ with the perturbative ansatz~\eqref{eq:eight_all_orders} for ${\rm O_{9}}$, by taking the limit $u_{4}, u_{8}, v_{4} \rightarrow 1$ of {\sc Cube 6789} in table~\ref{tab:octoriginsurfacesrelation}. The two expressions are seen to match perfectly. The associated coefficients $d_{i}$ are given to all loops by
\beq
\begin{aligned}
d_{1} &= -\frac{1}{96}(\Gamma'_{\pi/4}+2\tilde{\Gamma}_{0}+4\tilde{\Gamma}_{\pi/3}+2\tilde{\Gamma}_{-})\, , \\
d_{2} &= -\frac{1}{48}(\Gamma'_{\pi/4}+2\tilde{\Gamma}_{-})\, , \\
d_{3} &= -\frac{1}{48}(2\tilde{\Gamma}_{0}-2\tilde{\Gamma}_{\pi/3}+\tilde{\Gamma}_{+}) \, , \\
d_{4} &= -\frac{1}{48}(2\tilde{\Gamma}_{0}-\Gamma'_{\pi/4}-2\tilde{\Gamma}_{\pi/3}+\tilde{\Gamma}_{-})\, , \\
d_{5} &= -\frac{1}{48}(2\tilde{\Gamma}_{\pi/3}-2\tilde{\Gamma}_{0}+\tilde{\Gamma}_{+})\, ,
\end{aligned}
\eeq
where to save space we defined
\beq
\tilde{\Gamma}_{\pm} = (1+3^{\pm 1/2})\tilde{\Gamma}_{\pi/12} +  (1-3^{\pm 1/2})\tilde{\Gamma}_{5\pi/12}\, .
\eeq
As a cross check, one may verify that the exact same coefficients are obtained by matching the TBA predictions for $\Sigma_{8} = (+, +, -)$ and $\Sigma_{8} = (+,-, +)$ onto their corresponding line and surface. The $(+, +, -)$ result may be compared with the expressions for {\sc Pentagon 234} and {\sc Pentagon 345} in table~\ref{tab:octoriginsurfacesrelation}, in the limit $u_{8}, v_{4}\rightarrow 1$, using the ancillary file {\tt octagon\_QL\_formula.txt}. The $(+,-, +)$ result may be matched with the formula for {\sc Cube 6789}, after flipping $u_{i} \leftrightarrow u_{8-i}, v_{1}\leftrightarrow v_{2}, v_{3}\leftrightarrow v_{4}$ and taking the limit $v_{3}\rightarrow 1$.

\section{Superline}\label{appx:superline}

At $n=8$, we need an extra solution to the TBA equations to describe the super-origin $\rm O_{1}$, or, better yet, the {\sc Superline 1} connecting the two dihedral images of $\rm O_{1}$,
\beq
u_{1, \ldots , 8} \rightarrow 0\, , \qquad v_{1} = 1-v_{2} = v_{3} = 1-v_{4}\, ,
\eeq
with $v_{1}\in [0, 1]$. In terms of the OPE parameters, the line corresponds to
\beq\label{eq:super-scaling}
\tau_{1, 3} \sim -\sigma_{1,3} \sim -\tfrac{1}{2} \sigma_{2} \sim \tfrac{1}{4}\varphi_{1} \sim -\tfrac{1}{4}\varphi_{3} \rightarrow + \infty\, ,
\eeq
with $\varphi_{2}$ and $\tau_{2}$ kept fixed. Here, the numerical coefficients indicate how the parameters scale with respect to one another, with e.g.~$|\varphi_{1,3}|$ going to infinity 4 times faster than $\tau_{1,3}$. At first sight, one may think that this scaling is described by two large $Y$ functions, $Y_{1,1}$ and $Y_{-1, 3}$, for the two large chemical potentials in eq.~\eqref{eq:super-scaling}. This naive reasoning is not entirely correct however, because of the large $\sigma_s$ limit.

In order to find the right TBA description, one may draw inspiration from the analysis of the regular octagons~\cite{Alday:2010vh}. The latter refers to a continuous family of cyclic-symmetric kinematics,
\beq\label{eq:diagonal}
u_{1},\ldots, u_{8} = u\, , \qquad v_{1},\ldots,v_{4} = 1/2\, ,
\eeq
labelled by the cross ratio $u\in (0, \infty)$. It intersects {\sc Superline 1} at its midpoint ($v_{1} = v_{2}$) when $u\rightarrow 0$. In the TBA setup, the cyclic kinematics is associated to a 1-parameter family of constant $Y$-function solutions, which can be constructed exactly for any $u$. The solution reads, in our notation,
\beq
Y_{0, 1} = \frac{\sqrt{2 Y_{-1,1}}}{1+Y_{-1, 1}} = \frac{u}{1-u}\, ,
\eeq
with $Y_{a, s} = Y_{-a, 4-s}$ and $Y_{1, 1} Y_{-1, 1} = Y_{1, 2}Y_{0, 1} = Y_{0,2} = 1$. One concludes from it that four $Y$ functions are sent to infinity in the limit $u\rightarrow 0$, namely,
\beq\label{eq:largeY}
Y_{1, 1}\, , Y_{1, 2}\, , Y_{-1, 2} \, ,  Y_{-1, 3}\, ,
\eeq
assuming $Y_{1,1}>1$ for definiteness. Our working assumption is that the same system of four large $Y$ functions drives the QL behavior of the amplitude away from the cyclic point, all along {\sc Superline 1}.

Allowing for a dependence on $\theta$ in the $Y$ functions~\eqref{eq:largeY} and going to Fourier space, one finds that the problem is controlled by the 4-dimensional square matrix
\beq
\begin{aligned}
&1-K_{8}(z) = \frac{1}{(1+z)(1+z^{2})} \\
&\times \left[\begin{array}{cccc} 1+z^{3} & -z^{3} & z & 0 \\ - 1 & 1+z^{3} & -z(1+z) & z^{2} \\ z^{2} & -z(1+z) & 1+z^{3} & -1 \\ 0 & z & -z^{3} & 1+z^{3}  \end{array}\right]\, ,
\end{aligned}
\eeq
with coefficients following directly from eq.~\eqref{eq:TBA-z}.
The string integrand $\cS_{8}(z)$ is defined as usual, 
\beq
\cS_{8}(z) =  \hat{I}(1/z) \cdot \left[1-K_{8}(z)\right]^{-1} \cdot \hat{I}(z)^{T} \, ,
\eeq
with $\hat{I} = (\hat{I}_{+, 1}, \hat{I}_{+,2}, \hat{I}_{-,2}, \hat{I}_{-, 3})$. Straightforward algebra yields
\beq
\cS_{8}(z) = \frac{z \mathcal{P}_{8}^{{\rm super}}(z)}{(1-z^{4})(1-z^{8})}\, ,
\eeq
where $\mathcal{P}_{8}^{{\rm super}}(z) = z^{10}\mathcal{P}_{8}^{{\rm super}}(1/z)$ is a polynomial of degree 10 in $z$ and is quadratic in the OPE parameters. One concludes that the singularities of $\cS_{8}(z)$ lie on the unit circle, at the eight roots of unity, $1-z^{8} = 0$, corresponding to $\alpha = \{0, \pm \pi/8, \pm \pi/4, \pm 3\pi/8, \pi/2\}$.

To eliminate the OPE parameters in the limit~\eqref{eq:super-scaling}, one may use
\beq
\begin{aligned}
&u_{3}v_{1} \approx  e^{\tau_{3}-\sigma_{3}+\varphi_{3}}\, , \,\,\, u_{2} \approx e^{-2\tau_{3}}\, , \,\,\, u_{1}u_{4}v_{2} \approx e^{\tau_{3}+\sigma_{3}+\varphi_{3}}\, , \\
&u_{4} \approx e^{\tau_{2}+\sigma_{2}-\varphi_{2}}\, ,  \,\,\,\,\,\, v_{2}/v_{1} = e^{-2\tau_{2}}\, , \,\,\,\,\,\, u_{8} \approx e^{\tau_{2}+\sigma_{2}+\varphi_{2}}\, , \\
&u_{7}v_{1} \approx e^{\tau_{1}-\sigma_{1}-\varphi_{1}}\, ,  \,\,\, u_{6} \approx e^{-2\tau_{1}}\, , \,\,\, u_{5}u_{8}v_{2} \approx e^{\tau_{1}+\sigma_{1}-\varphi_{1}}\, ,
\end{aligned}
\eeq
where all quantities are small, except $v_{2}/v_{1} = O(1)$. Plugging these relations inside $\cS_{8}(z)$ and evaluating the master integral by residues around the unit-circle poles, one obtains the all-order remainder function on {\sc Superline 1} as a finite sum over angles. It reads,
\beq
\begin{aligned}
R_{8}  =  &-\sum_{\alpha}\frac{ \tilde{\Gamma}_{\alpha}\cos{\alpha}\cos{(3\alpha)}}{16\cos{(2\alpha)}} \bigg|\sum_{j=1}^{8} e^{2ij\alpha} l_{j}\bigg|^2 \\
&-\frac{\tilde{\Gamma}_{0}}{32} (\sum_{j=1}^{8}l_{j}+2\ell_{1}+2\ell_{2})^2-\frac{\tilde{\Gamma}_{\pi/2}}{8} (\ell_{1}-\ell_{2})^2 \, ,
\end{aligned}
\eeq
with the sum running over $\alpha = \{\pi/8, \pi/4, 3\pi/8\}$ and with $l_{i} = \ln{u_{i}}, \ell_{i} = \ln{v_{i}}$. Notice that only one term remains in the cyclic limit~\eqref{eq:diagonal}, namely, the one scaling with $\tilde{\Gamma}_{0} = \Gamma_{\textrm{oct}}-\Gamma_{\textrm{cusp}}$. The same happens at $n=6$, $R_{6} \sim -\tilde{\Gamma}_{0} \ln^{2}{(u_{1}u_{2}u_{3})}/24$, when approaching the origin along the diagonal $u_{1} = u_{2} = u_{3}$. It can be traced back to the fact that these limits are controlled by constant $Y$-function solutions.

Alternatively, we may expand the result over the basis of QL polynomials constructed with the baby amplitude bootstrap. The results match perfectly, with the coefficients
%
\begin{widetext}
\beq
\begin{aligned}
f_{1} &= -\frac{1}{64}(\Gamma'_{\pi/4}+2\tilde{\Gamma}_{0} + 2\tilde{\Gamma}_{\pi/8}+2\tilde{\Gamma}_{3\pi/8})\, ,
\qquad&f_{2} &= -\frac{1}{8} (\tilde{\Gamma}_{0}+\tilde{\Gamma}_{\pi/2})\, , \\
f_{3} &= -\frac{1}{32} (2\tilde{\Gamma}_{0}+\sqrt{2}\tilde{\Gamma}_{\pi/8}-\sqrt{2}\tilde{\Gamma}_{3\pi/8})\, ,
\qquad&f_{4} &= -\frac{1}{32} (2\tilde{\Gamma}_{0}-\Gamma'_{\pi/4})\, , \\
f_{5} &=  -\frac{1}{32} (2\tilde{\Gamma}_{0}-\sqrt{2}\tilde{\Gamma}_{\pi/8}+\sqrt{2}\tilde{\Gamma}_{3\pi/8})\, .
\end{aligned}
\eeq
\end{widetext}
We provide their explicit expressions at weak coupling
through 8 loops in the ancillary file {\tt octagon\_QL\_coefs.txt}.


\bibliography{biblio_Ostory}
\end{document}